\documentclass[a4paper,11pt]{article}
\pdfoutput=1 

\usepackage{jinstpub} 

\title{\boldmath Electromagnetic Calorimeter for MPD Spectrometer \\at NICA Collider}

\author[a,1]{A.Yu. Semenov,\note{Corresponding author.}}
\author[a]{S.~Bazylev,}
\author[a]{E.~Belyaeva,}
\author[a,b]{M. Bhattacharjee,}
\author[a,c]{B.~Dabrowska,}
\author[a]{D.~Egorov,}
\author[a]{V.~Golovatyuk,}
\author[a]{Yu.~Krechetov,}
\author[a]{A.~Shutov,}
\author[a]{V.~Shutov,}
\author[a]{S.~Sukhovarov,}
\author[a]{A.~Terletskiy,}
\author[a]{and I.~Tyapkin}

\affiliation[a]{Joint Institute for Nuclear Research, 141980, Dubna, Russia}
\affiliation[b]{Gauhati University, Guwahati, Assam 781014, India}
\affiliation[c]{Plovdiv University "Paisii Hilendarski", Tzar Assen 24, Plovdiv, Bulgaria}

\emailAdd{semenov\underline{~}andrei@yahoo.com}

\abstract{The Multi-Purpose Detector (MPD) is designed to study a hot and dense baryonic matter formed in heavy-ion collisions at $\sqrt{s_{NN}}$=4-11 GeV at the NICA accelerator complex (Dubna, Russia). Large-sized electromagnetic calorimeter (ECal) of the MPD spectrometer will provide precise spatial and energy measurements for photons and electrons in the central pseudorapidity region of |$\eta$|<1.2. The Shashlyk-type sampling structure of the ECal is optimized for the photons energy range from about 40 MeV to 2-3 GeV. Fine segmentation and projective geometry of the calorimeter allow to deal with high multiplicity of secondary particles from Au-Au reactions. In this talk, we report on a design, a construction status and expected parameters of the ECal.}


\collaboration[c]{on behalf of MPD collaboration}

\proceeding{Calorimetry for High Energy Frontier Conference (CHEF 2019)\\
  November 25-29, 2019\\
  Fukuoka, Japan}

\begin{document}
\maketitle
\flushbottom

\section{Introduction}
\label{sec:intro}
\begin{figure}[htbp]
\centering 
\includegraphics[width=.45\textwidth]{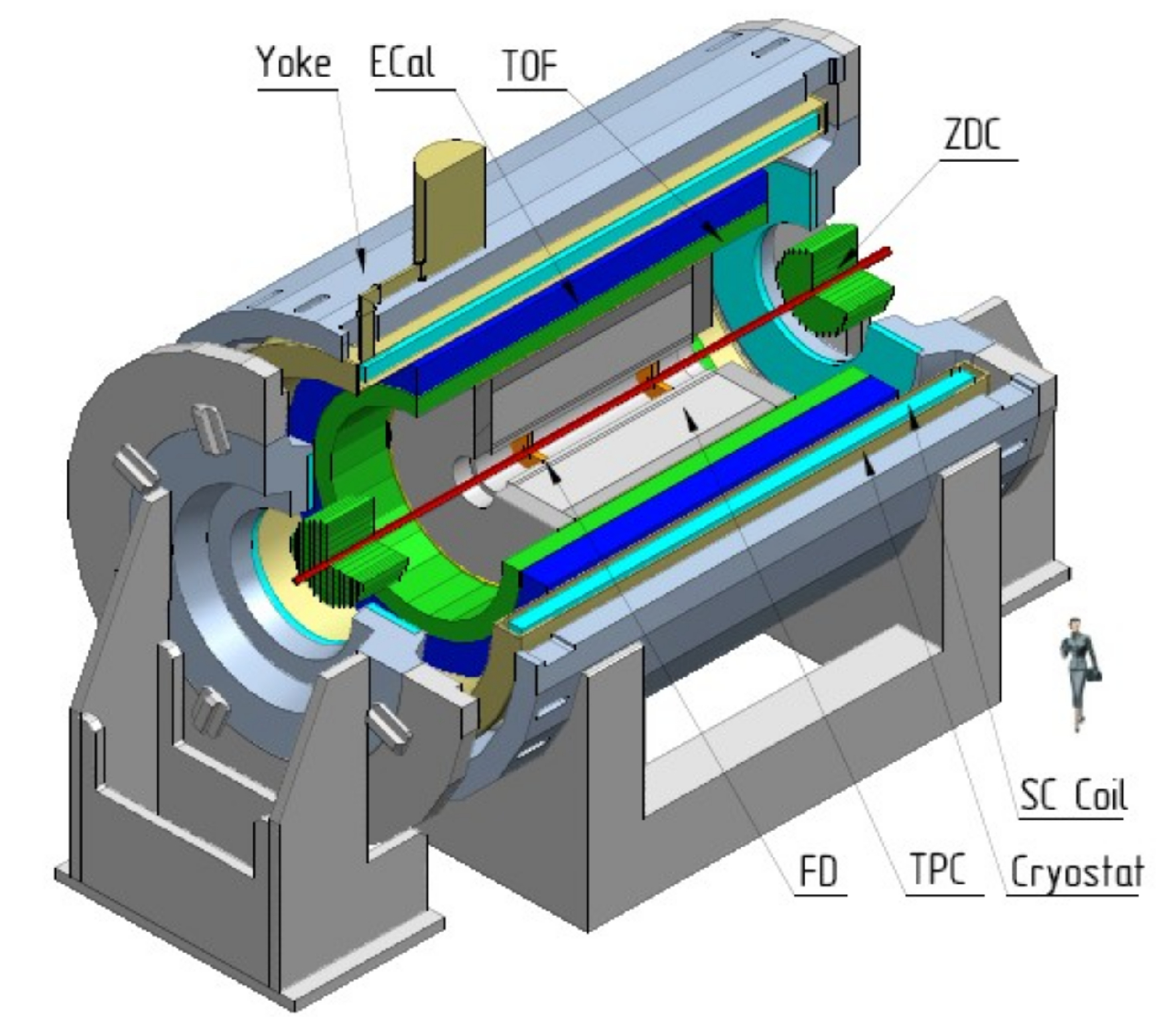}
\caption{\label{fig:1} MPD setup. The ECal location is shown in blue.}
\end{figure}

The main goal of the Multi-Purpose Detector (MPD) at  Nuclotron-based Ion Collider fAcility (NICA) is a study of signals from hot baryonic matter with maximal density that might be produced in collisions of heavy ions  at $\sqrt{s_{NN}}$=4-11 GeV \cite{mpd,11}. The experimental program with heavy ions at NICA includes event-by-event measurements of observables that are expected to be sensitive to high-density effects and phase transitions: particle yields, particle yields ratios, fluctuations, and correlations. The MPD spectrometer (Fig.~\ref{fig:1}) has 2-$\pi$ acceptance in azimuth, low material budget, and is able to handle event rate up to 6~kHz. Solenoid magnet with 0.5~T magnetic field and the barrel Time-Projection-Chamber (TPC) tracking system provide the measurements of momenta of charged particles with sufficient accuracy (about 2\% at $p_t$=300~MeV/c), and vertex finding. $dE/dx$ measurements with TPC together with Time-of-Flight measurements allow $\pi/K$ separation up to 1.5~GeV/c, and $K/p$ separation up to 3 GeV/c.
Electromagnetic barrel calorimeter (ECal) as an important part of MPD provides an access to the electromagnetic probes such as direct photons and lepton pairs, decays of neutral mesons, and significantly improves electron/hadron separation.

\section{Electromagnetic Calorimeter Modules}
\label{sec:calo}
Large-sized (about 6-meters-long and 4.5-meters in diameter) ECal covers the central pseudorapidity region of |$\eta$|<1.2 (Fig. 1), and is optimized for precise spatial and energy measurements for photons and electrons in the energy range from about 40 MeV to 2-3 GeV \cite{tdr}. To deal with a high multiplicity of secondary particles from central heavy-ions collisions, ECal has a fine segmentation and consists of 38,400 cells (towers). Taking all requirements (high energy resolution, large enough distance to the vertex, small Moliere radius, ability to work in the high magnetic field, high time resolution, and a reasonable price) into consideration, a "shashlyk"-type electromagnetic calorimeter was selected \cite{shashlyk}.
Each tower has a sandwich structure of 210 polystyrene scintillator \cite{plastic1,plastic2} and 210 lead plates with 16 Wave Length Shifting (WLS) fibers Kuraray Y-11~(200) \cite{kuraray} that penetrate the plates to collect the scintillation light and transport it to the photodetector; the far (from the photodetector) end of fiber is painted with white light-reflecting paint to increase an amount of the collected light.
The thickness of each scintillator plate is 1.5 mm, and the thickness of lead plate is 0.3~mm. Monte-Carlo simulations suggest that such a proportion of scintillators and lead coverters provides the sampling fraction of about 34-39\% (depending on energy), and results into relatively small statistical term and a good energy resolution in the energy range below 1~GeV. A dark side of this calorimeter design is that the limited space inside the MPD magnet leads to the limited ECal thickness just above 11~X$_0$ and the correspondent visible energy leak from the backend of the calorimeter; though the leak does not exceed 10-12\% in the ECal energy range.  

\begin{figure}[htbp]
\centering 
\includegraphics[width=.8\textwidth]{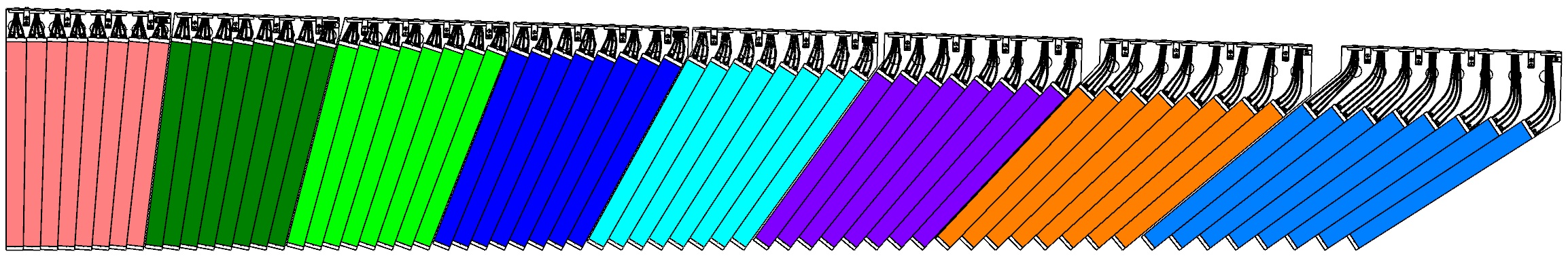}\\[2mm]
\qquad
\hspace*{-7mm}\includegraphics[width=.21\textwidth,origin=a]{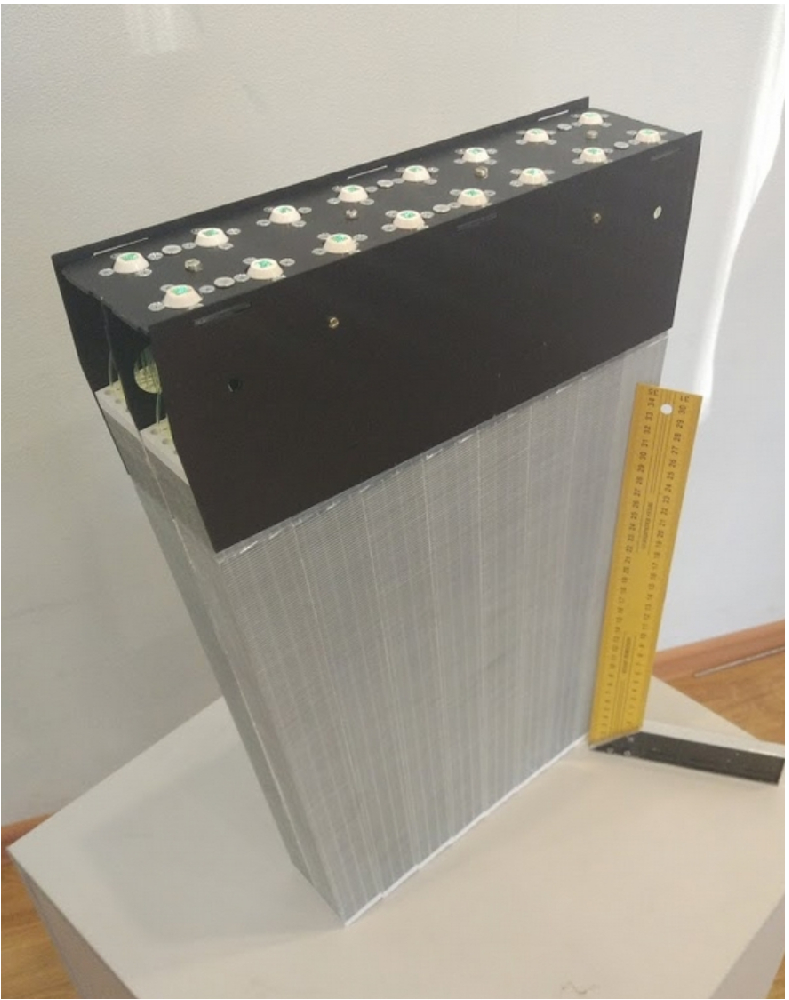}
\hspace*{2mm}\includegraphics[width=.58\textwidth,origin=b]{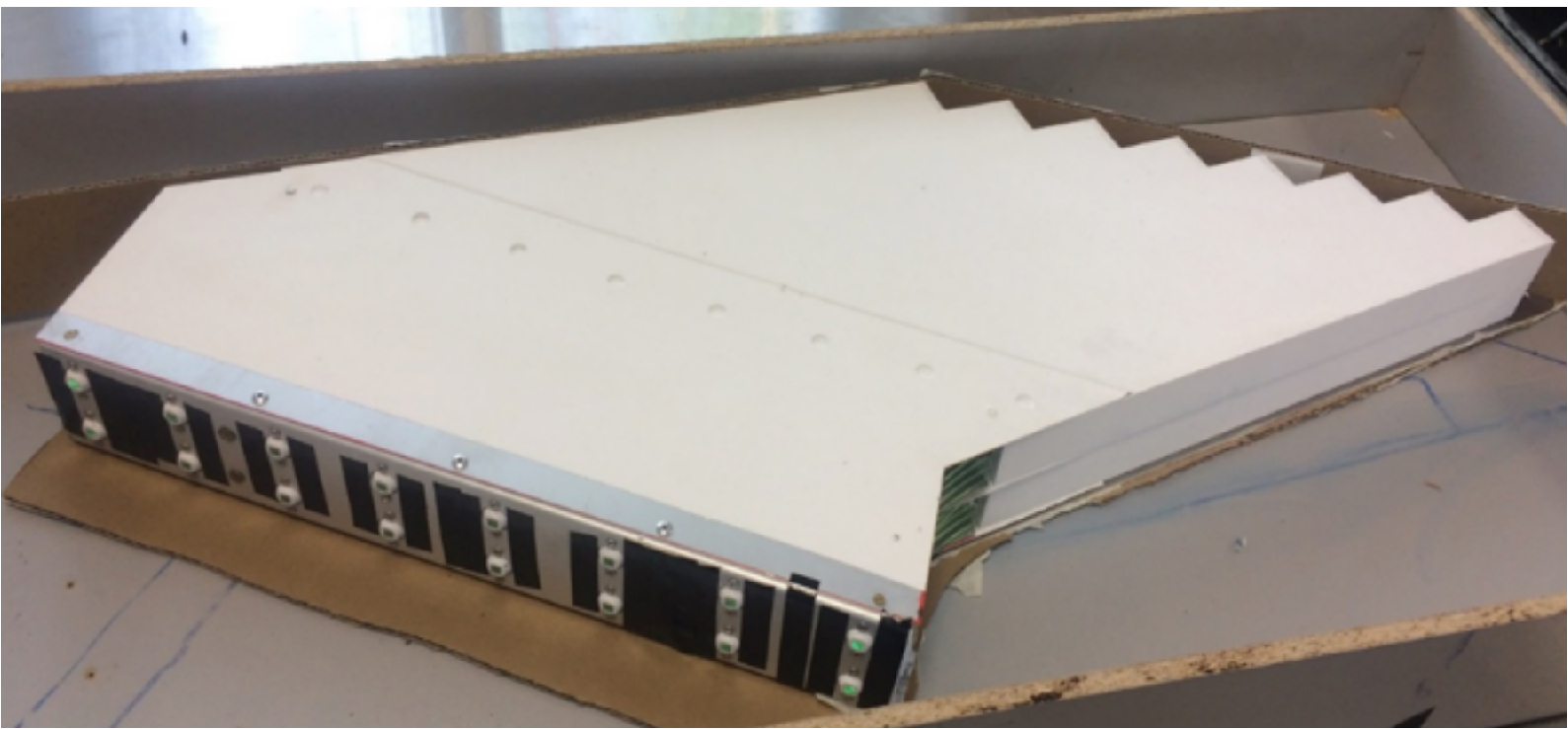}
\caption{\label{fig:2} Top: evolution of ECal modules shape along the beam direction from the center to the edge of MPD. Bottom left: photo of the module from the MPD center (type 0 shown in pink in the top panel). Bottom right: photo of the module from the edge of MPD (type 7 shown in medium-blue in the top panel).}
\end{figure}
Each ECal module consists of 16 towers that are glued together. The geometry of the each module depends on the module Z-coordinate (beam direction) location in respect to the beams interception point (Fig.~\ref{fig:2}).
The advantage  the calorimeter with the projective geometry (where the towers are inclined along the beam axis to keep the tower axis to be consistent with the direct view to the beams intersection region) is a reduction of dead zones, increase of the detector efficiency, improvement of a linearity and an energy resolution of the calorimeter measurements in conditions of high multiplicity of secondary particles from the collisions of heavy ions. 

In total, ECal will contain 2,400 modules of 8 different types. The production of the ECal modules is divided between Russian (25\%) and Chinese (75\%) facilities. Production of the modules in Russia is started in 2019, and the production in China is expected to be started in 2020. 

\section{ECal Power Structure}
From geometrical point of view, ECal is divided in 25 sectors or 50 half-sectors; each half-sector contains 6$\times$8=48 modules of 8 different types. These modules are located in the half-sector container (basket) made of fiberglass material. Rigidity of the container is enough to provide deformation less than 0.5 mm under full half- sector load of about 1.5 tons. 

\begin{figure}[htbp]
\centering 
\qquad
\hspace*{-7mm}\includegraphics[width=.43\textwidth,origin=a]{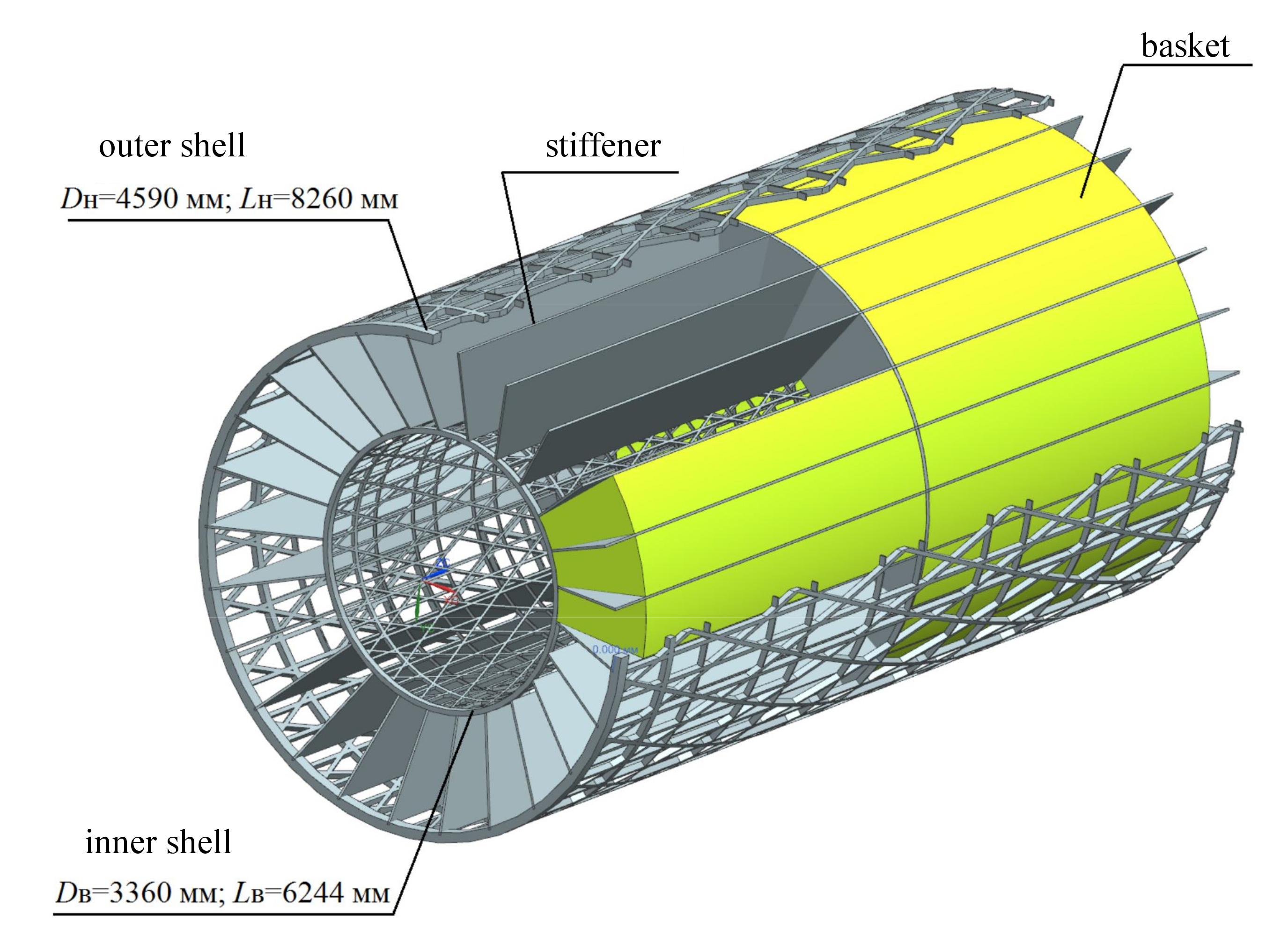}
\hspace*{2mm}\includegraphics[width=.36\textwidth,origin=b]{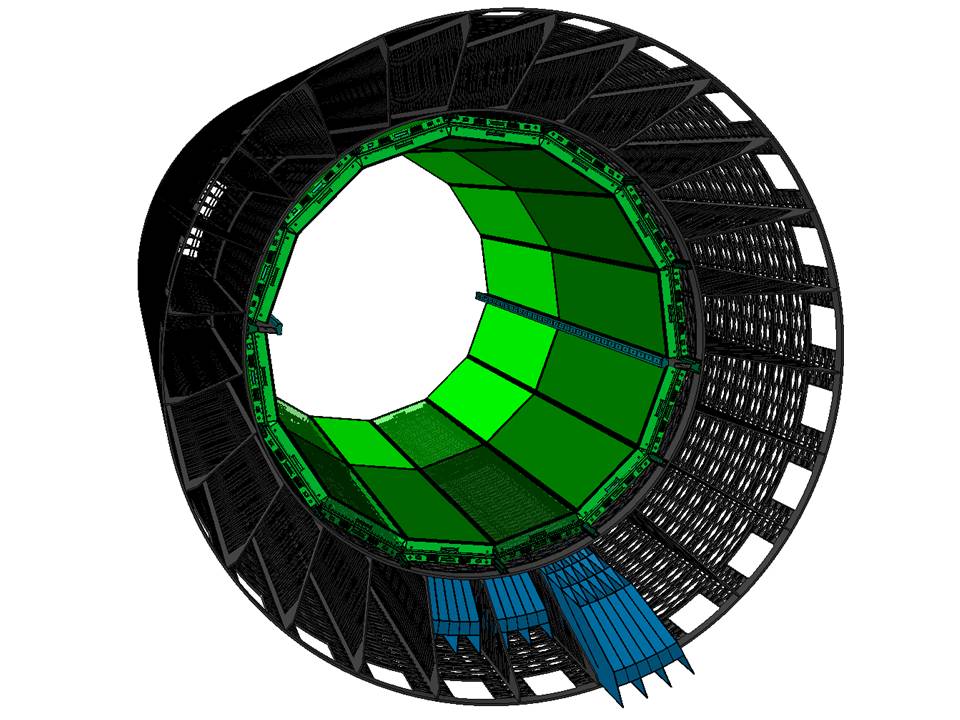}
\caption{\label{fig:3} The power frame of ECal. On the right panel, the half-sector containers (baskets) are shown in blue, and Time-of-Flight modules are shown in green.}
\end{figure}
Original MPD construction plan was to build ECal as a self- supporting structure. After sharing the modules production between Russian and Chinese institutions and corresponding differences in the modules delivery schedule, the decision was made to locate the calorimeter half-sectors into carbon-composite power MPD frame which is strong enough to hold the total weight of ECal and other MPD detectors (about 100 tons) with a maximal deformation of 2-3 mm, that makes possible to install and extract any ECal half-sector without dismounting whole MPD spectrometer.

In addition to the modules, half-sector includes ECal readout electronics. To keep ability to extract and reinstall calorimeter electronics (for service and repair), special electronics installation system was developed. Electronics support is provided by 3-m-long boxes; each box serves 2$\times$8=16 modules. Outside each box, 16 front-end boards with 16 photodetectors (6$\times$6~mm$^2$ Hamamatsu S13360-6025PE MAPD \cite{sipm1,sipm2}), preamplifiers and slow-control electronics (that controls the temperature and makes corresponding correction of about 45~mV/deg. on photodetectors supply voltage) are located, while the JINR-designed 64-channel 14-bit 62.5~MS/s Pipelined ADC64ECAL boards \cite{ADC} are housed inside heat-isolated box to minimize influence on photodetectors. The hit production is estimated as about 150~W per half-sector (or about 7.5~kW for whole ECal), and the water-cooling system is used to evacuate the heat from the boxes. 
\begin{figure}[htbp]
\centering 
\qquad
\includegraphics[height=.3\textwidth,origin=a]{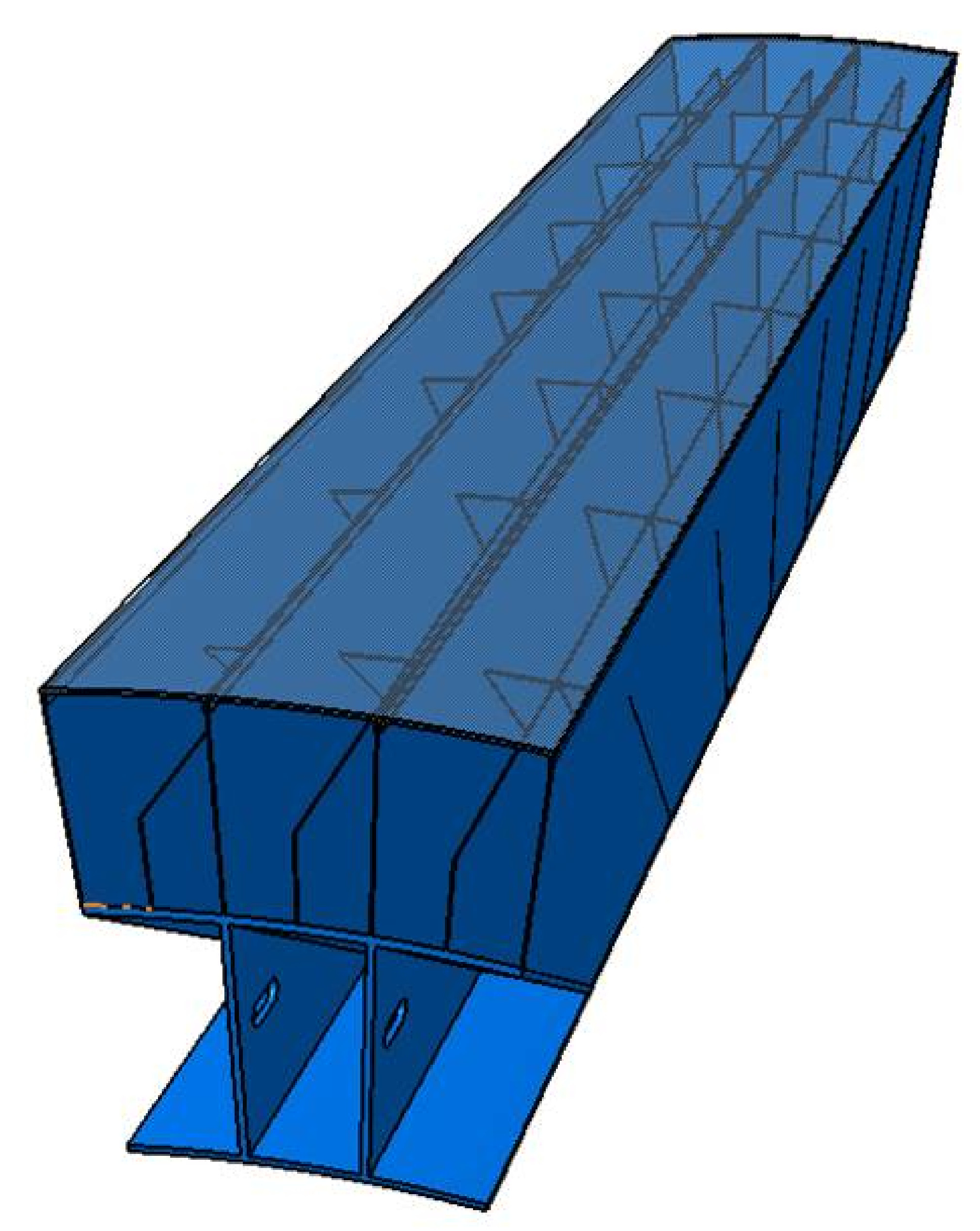}
\hspace*{15mm}\includegraphics[height=.3\textwidth,origin=b]{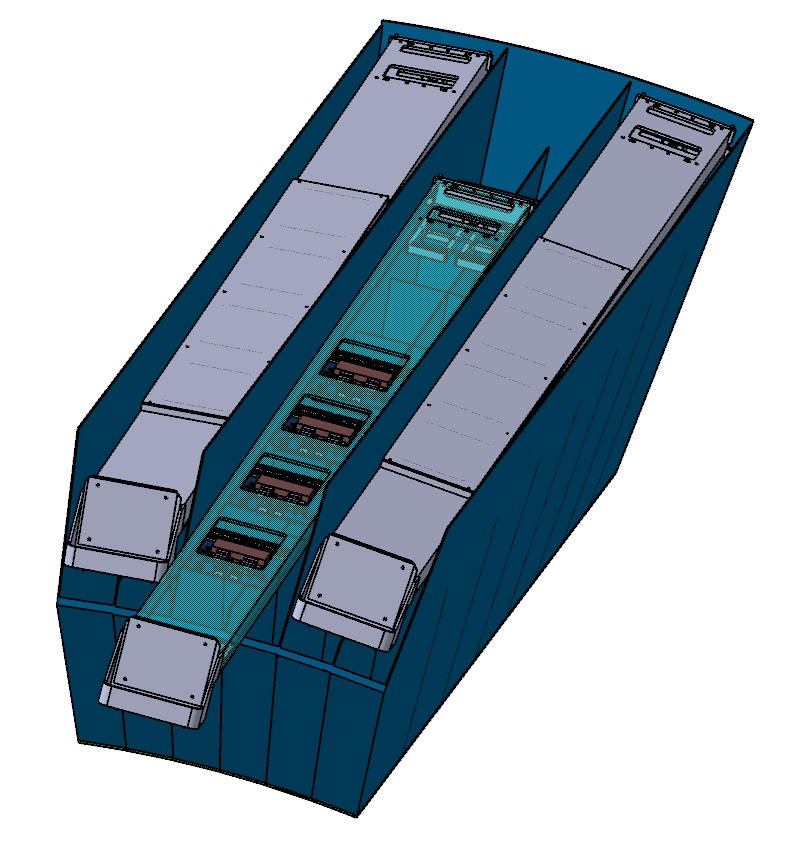}
\caption{\label{fig:4} Left panel shows the container for ECal half-sector (basket). Right panel shows the installation of boxes with electronics into the baskets.}
\end{figure}

\section{ECal Module Tests}
\begin{figure}[htbp]
\centering 
\qquad
\includegraphics[height=.24\textwidth,origin=a]{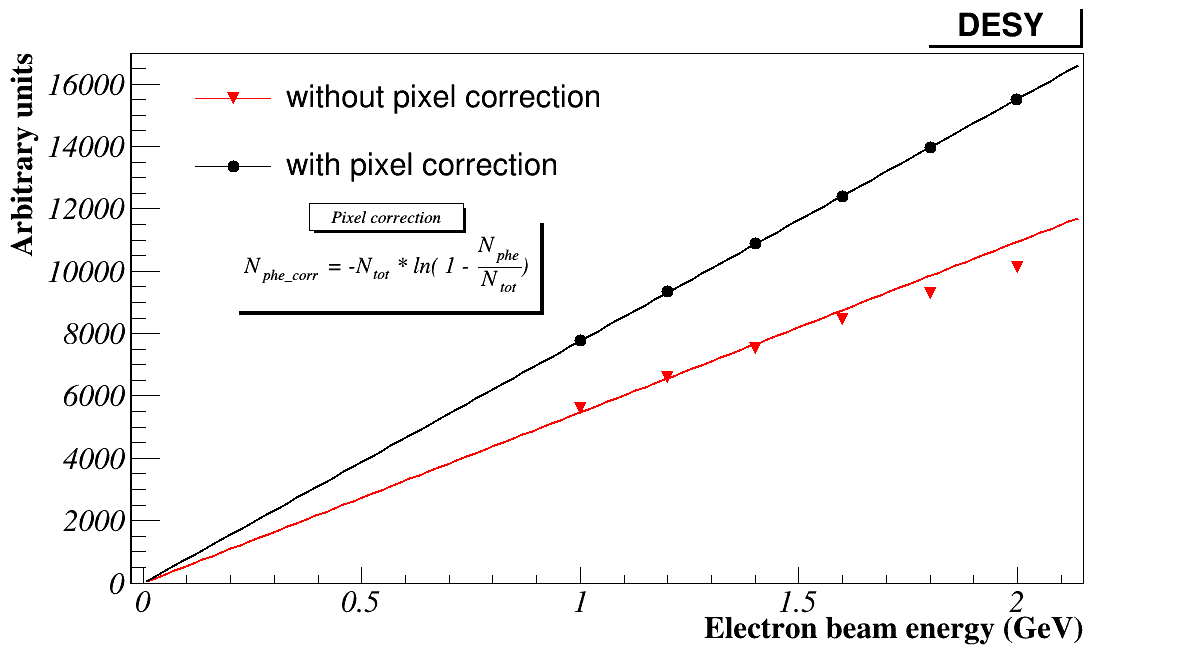}
\includegraphics[height=.24\textwidth,origin=b]{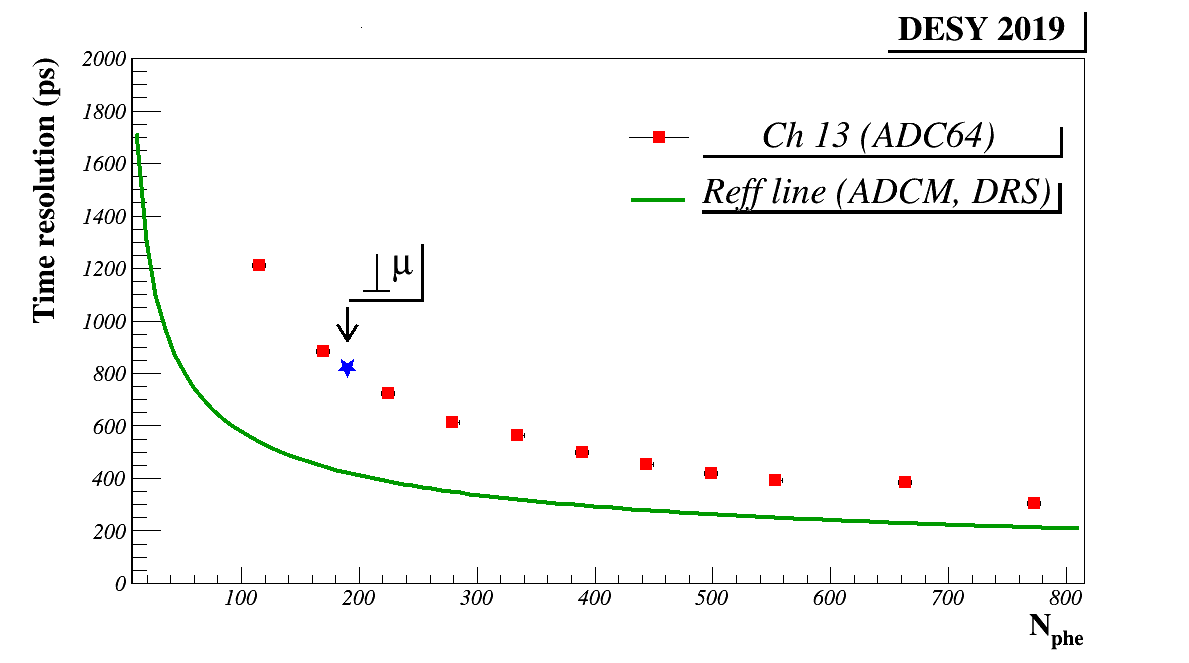}
\caption{\label{fig:5} Results of the linearity (left panel) and the time resolution (right panel) measurements of prototype ECal modules with the electron beam at GSI.}
\end{figure}

Tests of the prototype modules were performed with electron beams at DESY (Hamburg, Germany) and Lebedev Physics Institute of Russian Academy of Science (Troitsk, Russia). For electron beam energies above 1 GeV, a visible deviation from linearity for the ECal response was observed (see left panel in Fig.~\ref{fig:5}). It was found that this deviation is connected mostly with the signal saturation because of the limited number of pixels in the MAPD; the correction on this effect restores the linearity. The signal time was produced from an analysis of ADC waveform front, and the time resolution for each tower is presented in Fig.~\ref{fig:5} (right panel). The blue star in this plot is belong to the time measurement with cosmic muons that travels through ECal modules in a transverse direction, and this result is consistent with measurements on electron beam. The green line on the plot presents results of a special time measurement with high-precision and high-frequency ADCs, and allows to estimate the  contribution of "standard" ECal electronics into measured time resolution.

\begin{figure}[htbp]
\centering 
\includegraphics[height=.25\textwidth]{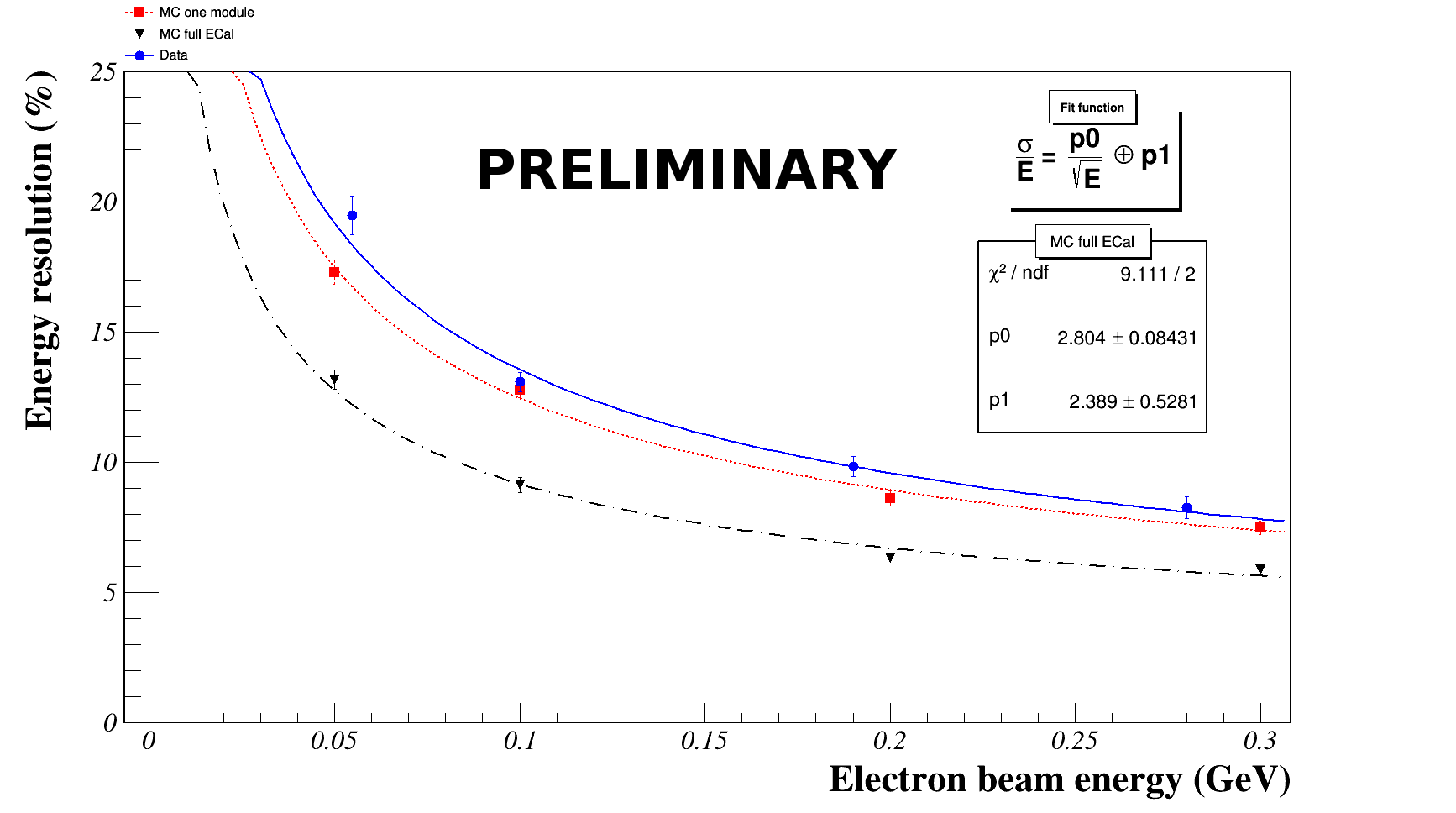}
\caption{\label{fig:6} Results of the energy resolution measurements with the electron beam in Troitsk.}
\end{figure}
An energy resolution of the single ECal module was measured recently with relatively-low-energy electron beam in Troitsk. The obtained data (shown in blue in Fig.~\ref{fig:6}) are in a good agreement with results of Monte-Carlo simulations for a single module (shown in red in Fig.~\ref{fig:6}). The same Monte-Carlo simulation made with whole calorimeter (dash-dotted black line in Fig.~\ref{fig:6}) allows to make a preliminary estimation of the expected energy resolution of ECal: 
\begin{equation}
\label{eq:x}
\Delta{E}/E \approx \frac{3.0\%}{\sqrt{E(GeV)}} \oplus 2.4\%
\end{equation}

\section{Conclusions}
Large-sized barrel electromagnetic calorimeter for MPD spectrometer is under construction in Joint Institute for Nuclear Research (Dubna, Russia). The "shashlyk"-type calorimeter is optimized to deal with high multiplicity of secondary particles from heavy-ion collisions at NICA accelerator complex. Production of the calorimeter modules is shared between Russian and Chinese institutions. The prototype modules tests with electron beams in GSI and Troitsk demonstrate that the measured calorimeter parameters are in a good agreement with expectations.

\acknowledgments

This work was supported by the Ministry of Education and Science of the Russian Federation under Contract No.05.615.21.0005 in the framework of
the project RFMEFI61518X0007.



\begin{thebibliography}{99}

\bibitem{mpd} V.D. Kekelidze, \emph{Phys. Part. Nucl.} {\bf 49} (2018) 457.

\bibitem{11} V. Golovatyuk, V. Kekelidze, V. Kolesnikov, et al., \emph{The Multi-Purpose Detector (MPD)
of the collider experiment}, \emph{The European Physical Journal} {\bf A 52 (8)} (2016) 212.

\bibitem{tdr} MPD NICA Technical Design Report of the Electromagnetic Calorimeter (ECal), rev. 3.8, September 2019. http://mpd.jinr.ru/doc/mpd-tdr/

\bibitem{shashlyk} G.S. Atoian et al., \emph{Nucl. Instrum. Meth.} {\bf A 584} (2008) 291.

\bibitem{plastic1} https://www.polipak.ru/

\bibitem{plastic2} http://uniplast-vladimir.com/scintillation

\bibitem{kuraray} http://kuraraypsf.jp/pdf/ws.html

\bibitem{sipm1} https://www.hamamatsu.com/us/en/product/type/S13360-6025PE/index.html

\bibitem{sipm2} https://www.hamamatsu.com/resources/pdf/ssd/s13360\underline{~~}series\underline{~~}kapd1052e.pdf

\bibitem{ADC} https://afi.jinr.ru/ADC64s2





\end{thebibliography}
\end{document}